\documentclass[pra,aps,twocolumn,preprintnumbers,superscriptaddress]{revtex4}

\usepackage{amsfonts,amssymb,amsmath,amsthm}
\usepackage{graphicx}
\usepackage{url}
\newcommand{\<}{\langle}
\renewcommand{\>}{\rangle}

\makeatletter
\def\revddots{\mathinner{\mkern1mu\raise\p@ 
\vbox{\kern7\p@\hbox{.}}\mkern2mu 
\raise4\p@\hbox{.}\mkern2mu\raise7\p@\hbox{.}\mkern1mu}} 
\makeatother

\newtheorem*{sdp}{Semidefinite Program}
\newtheorem{lemma}{Lemma}
\newtheorem{theorem}{Theorem}

\newcommand{\Tr}{\mathrm{Tr}}

\newcommand{\CC}{{\mathbb{C}}}
\newcommand{\RR}{{\mathbb{R}}}
\newcommand{\ZZ}{{\mathbb{Z}}}

\newcommand{\eg}{\textit{e.g.}}
\newcommand{\etal}{\textit{et~al.}}
\newcommand{\ie}{\textit{i.e.}}

\begin{document}

\title{Improved quantum algorithms for the ordered \\
       search problem via semidefinite programming}

\author{Andrew M. \surname{Childs}}
\email[]{amchilds@caltech.edu}
\affiliation{Institute for Quantum Information,
             California Institute of Technology,
             Pasadena, CA 91125, USA}

\author{Andrew J. \surname{Landahl}}
\email[]{alandahl@unm.edu}
\affiliation{Center for Advanced Studies,
             Dept.\ of Physics and Astronomy,
             University of New Mexico,
             Albuquerque, NM, 87131, USA}

\author{Pablo A. \surname{Parrilo}}
\email[]{parrilo@mit.edu}
\affiliation{Laboratory for Information and Decision Systems,
             Massachusetts Institute of Technology,
	         Cambridge, MA 02139, USA}

\date[]{21 August 2006}


\begin{abstract}
One of the most basic computational problems is the task of finding a
desired item in an ordered list of $N$ items.  While the best
classical algorithm for this problem uses $\log_2 N$ queries to the
list, a quantum computer can solve the problem using a constant factor
fewer queries.  However, the precise value of this constant is
unknown.  By characterizing a class of quantum query algorithms for
ordered search in terms of a semidefinite program, we find new quantum
algorithms for small instances of the ordered search problem.
Extending these algorithms to arbitrarily large instances using
recursion, we show that there is an exact quantum ordered search algorithm
using $4 \log_{605} N \approx 0.433 \log_2 N$ queries, which improves
upon the previously best known exact algorithm.
\end{abstract}

\pacs{03.67.Lx}

\maketitle

\section{Introduction}

The \emph{ordered search problem} (OSP) is the problem of finding the first
occurrence of a target item in an ordered list of $N$ items subject to the
promise that the target item is somewhere in the list.  Equivalently, we can
remove the promise by viewing the OSP as the problem of finding the earliest
insertion point for a target item in a sorted list of $N-1$ items.  The OSP
is ubiquitous in computation, not only in its own right, but also as a
subroutine in algorithms for related problems, such as sorting.

To characterize the computational difficulty of the ordered search problem,
we are interested in knowing how many times the list must be queried to find
the location of the target item.  The minimal number of queries required to
solve the problem in the worst case is known as its \emph{query complexity}.
Using information theoretic arguments, one can prove that any deterministic
classical algorithm for the OSP requires $\lceil\log_2 N\rceil$ queries.
This lower bound is achieved by the well-known binary search algorithm
\cite{Knuth:1998a}.

Quantum computers can solve the ordered search problem using a number of queries that is smaller by a constant factor than the number of queries used in the binary search algorithm.  The best known lower bound, proved by H{\o}yer, Neerbek, and Shi, shows that any quantum algorithm for the OSP that is \emph{exact} (\ie, succeeds with unit probability after a fixed number of queries) requires at least $(\ln N - 1)/\pi \approx 0.221 \log_2 N$ queries  \cite{Hoyer:2002a}.  In other words, at most a constant factor speedup is possible.  The best published exact quantum OSP algorithm, obtained by Farhi, Goldstone, Gutmann, and Sipser, uses $3\lceil\log_{52} N\rceil \approx 0.526 \log_2 N$ queries, showing that a constant factor speedup is indeed possible \cite{Farhi:1999b}.  However, there remains a gap between the constants in these lower and upper bounds.  Since the OSP is such a basic problem, it is desirable to establish the precise value of the constant factor speedup for the best possible quantum algorithm: this constant is a fundamental piece of information about the computational power of quantum mechanics.

In this article, we study the query complexity of the ordered search problem
by exploiting a connection between quantum query problems and convex
optimization.  Specifically, we show that the existence of an algorithm for
the OSP that is \emph{translation invariant} (in the sense of
\cite{Farhi:1999b}) is equivalent to the existence of a solution for a
certain semidefinite program (SDP).  By solving this semidefinite program
numerically, we show that there is an exact quantum query algorithm to
search a list of size $N=605$ using $4$ queries.

Since the size of the semidefinite program increases as we increase $N$, we
cannot directly perform a numerical search for a quantum ordered search
algorithm for arbitrarily large problem instances.  However, by applying the
$4$-query algorithm recursively, we see that there is an exact algorithm for a
list of size $N$ using $4 \log_{605} N \approx 0.433 \log_2 N$ queries.
Thus, our result narrows the gap between the best known algorithm and the
lower bound of \cite{Hoyer:2002a}.  In particular, this shows that the
quantum query complexity of the OSP is strictly less than $\log_2 \sqrt{N}$,
which one might have naively guessed was the query complexity of ordered
search by analogy with the \emph{unordered search problem}, whose quantum
query complexity is $\Theta(\sqrt{N})$ \cite{Grover:1996a,Bennett:1997a}.

In addition to providing a way of searching for algorithms, the semidefinite programming approach has the advantage that a solution to the dual SDP provides a certificate of the non-existence of an algorithm.  Thus we are able to provide some evidence (although not a proof) that $N=605$ is the largest size of a list that can be searched with $k=4$ queries, by showing that no algorithm exists for $N=606$.

The remainder of the article is organized as follows.  In
Section~\ref{sec:invariant}, we describe the class of translation invariant
algorithms that we focus on and summarize known results about such
algorithms.  In Section~\ref{sec:semidefinite}, we show how these algorithms
can be characterized as the solutions of a semidefinite program.  Finally,
in Section~\ref{sec:results}, we present the results obtained by solving
this semidefinite program, and conclude with a brief discussion.

\section{Translation invariant quantum algorithms for ordered search}
\label{sec:invariant}

\subsection{Query models for ordered search}

In the standard query model for the ordered search problem, sometimes known
as the \emph{comparison model}, a query to the $x$th position of the list
indicates whether the target item occurs before or after (or at) that
position.  If the target item is at position $j\in\{0,1,\ldots,N-1\}$, then
its location is encoded in the function $f_j: \{0,1,\ldots,N-1\} \to \{\pm
1\}$ defined as
\begin{align}
  f_j(x) &:= \begin{cases}
         -1 & x < j \\
         +1 & x \geq j.
             \end{cases}
\end{align}
When searching an explicit list with no information about its structure
other than the fact that it is ordered, this function captures essentially
all the information that is available from examining a given position
in the list.  The ordered search problem is to determine $j$ using as
few queries to $f_j$ as possible.

The ordered search problem has a kind of symmetry: for $j \in
\{0,1,\ldots,N-2\}$, if we change the target item from $j$ to $j+1$, then we
find
\begin{align}
  f_{j+1}(x) = \begin{cases}
                 -1       & x=0 \\
                 f_j(x-1) & 1 \le x < N.
               \end{cases}
\label{eq:finvariance}
\end{align}
Unfortunately, we must handle what happens at the boundary (namely,
at $x=0$) as a special case.  However, as observed in \cite{Farhi:1999b},
we can remedy the situation by extending $f_j$ to the function
$g_j: \ZZ/2N \to \{\pm 1\}$ defined as
\begin{align}
  g_j(x) &:= \begin{cases}
                f_j(x)   & 0 \leq x < N \\
               -f_j(x-N) & N \leq x < 2N
             \end{cases}
\label{eq:gquery}
\end{align}
for $j \in \{0,1,\ldots,N-1\}$, and
\begin{align}
  g_j(x) &:= -g_{j-N}(x)
\end{align}
for $j \in \{N,N+1,\ldots,2N-1\}$,
where all arithmetic is done in $\ZZ/2N$, \ie, modulo $2N$.  The advantage
of using this modified function is that the symmetry expressed in
(\ref{eq:finvariance}) now appears without special boundary conditions
as a \emph{translation equivariance} in the group $\ZZ/2N$, namely as
\begin{align}
  g_{j+\ell}(x) &= g_{j}(x-\ell) \qquad \forall\, j,x,\ell \in \ZZ/2N.
\label{eq:ginvariance}
\end{align}

Although the functions $g_j$ are defined for all $j \in \ZZ/2N$, it is sufficient to consider the problem of determining $j$ with the promise that $j \in \{0,1,\ldots,N-1\}$.  (Indeed, with the quantum phase oracle for $g_j$ that we define in the next section, it will turn out that $j$ is indistinguishable from $j+N$.)  For this problem, the functions $f_j$ and $g_j$ are equivalent, in the sense that any algorithm using one type of query can be mapped onto an algorithm using the same number of the other type of query.  A single query to $f_j$ can be simulated by simply querying $g_j$ on the original value of $x \in \{0,1,\ldots,N-1\}$.  On the other hand, one query to $g_j$ can be simulated by a query to $f_j$ pre- and post-processed according to (\ref{eq:gquery}).  Thus, there is no loss of generality in using the modified function $g_j$ instead of the original function $f_j$: the query complexity of the OSP is the same using either type of query.

\subsection{Quantum query algorithms}

In the quantum mechanical version of the query model, access to the query
function is provided by a unitary transformation.  Specifically, we will use
the \emph{phase oracle} for $g_j$, a linear operator $G_j$ defined by the
following action on the computational basis states $\{|x\>: x \in
\ZZ/2N\}$:
\begin{align}
  G_j |x\> &:= g_j(x) |x\>.
\end{align}
A $k$-query quantum algorithm is specified by an initial quantum
state $|\psi_0\>$ and a sequence of ($j$-independent) unitary operators
$U_1,U_2,\ldots,U_k$.  The algorithm begins with the quantum computer in
the state $|\psi_0\>$, and query transformations and the operations $U_j$
are applied alternately, giving the final quantum state
\begin{align}
  |\phi_j\> &:= U_k G_j U_{k-1}\ldots U_1 G_j|\psi_0\>.
\end{align}
We say the algorithm is \emph{exact} if
$\<\phi_j|\phi_{j'}\>=\delta_{j,j'}$ for all $j,j' \in
\{0,1,\ldots,N-1\}$, since in this case there is some measurement that
can determine the result $j \bmod N$ with certainty.
(Note that the final unitary $U_k$ has no effect on whether the algorithm
is exact, but it is convenient to include for the purposes of the
following discussion.)
For each value of $N$, our goal is to find choices of $|\psi_0\>$ and
$U_1,U_2,\ldots,U_k$ for $k$ as small as possible so that the resulting
quantum algorithm is exact.

The search for a good quantum algorithm for the OSP can be considerably
simplified by exploiting the translation equivariance (\ref{eq:ginvariance})
of the function $g_j$ \cite{Farhi:1999b}.  This equivariance manifests itself as a symmetry of the query operators.  In terms of the translation operator $T$ defined by
\begin{align}
  &T |x\>    := |x+1\> \quad \forall\, x \in \ZZ/2N
\end{align}
(where addition is again performed in $\ZZ/2N$),
we have
\begin{align}
  T G_j T^{-1} &= G_{j+1}
  \quad \forall\, j \in \ZZ/2N.
\label{eq:QMequivariance}
\end{align}
Thus, it is natural to choose the quantum algorithm to have the
translation invariant initial state
\begin{align}
  |\psi_0\> = \frac{1}{\sqrt{2N}} \sum_{x=0}^{2N-1} |x\>
\label{eq:unifstate}
\end{align}
satisfying $T |\psi_0\> = |\psi_0\>$, and translation invariant unitary
operations $U_t$, \ie, unitary operators satisfying
\begin{align}
  T U_t T^{-1} = U_t
\label{eq:tinvunitary}
\end{align}
for $t \in \{1,2,\ldots,k\}$. 
Of course, while (\ref{eq:QMequivariance}) holds for all $j \in \ZZ/2N$, we are promised that $j \in \{0,1,\ldots,N-1\}$.  Correspondingly, we can require the $N$ possible orthogonal final states to label the location of the marked item as follows:
\begin{align}
  |\phi_j\> &:= \begin{cases}
    \frac{1}{\sqrt{2}}\left(|j\> + |j + N\>\right) & \text{$k$ even} \\
    \frac{1}{\sqrt{2}}\left(|j\> - |j + N\>\right) & \text{$k$ odd}
  \end{cases}
\label{eq:finalstates}
\end{align}
(where the separation into $k$ even and odd is done for reasons explained in \cite{Farhi:1999b}).
Overall, we refer to an algorithm with the initial state (\ref{eq:unifstate}), unitary operations satisfying (\ref{eq:tinvunitary}), and the final states (\ref{eq:finalstates}) as an \emph{exact, translation invariant algorithm} (in the sense of
\cite{Farhi:1999b}).

An advantage of a translation invariant algorithm for ordered search is
that, if it can find the target item when $j=0$, then it can find the
target item for all values of $j$.  Using (\ref{eq:QMequivariance}),
we have $T^{-j} G_j T^j = G_0$.  Thus
\begin{align}
  |\phi_j\> &= (T^j U_k T^{-j}) G_j \ldots
               (T^j U_1 T^{-j}) G_j (T^j |\psi_0\>) \\
            &= T^j U_k (T^{-j} G_j T^j) U_{k-1} \ldots
                   U_1 (T^{-j} G_j T^j) |\psi_0\> \\
            &= T^j U_k G_0 U_{k-1} \ldots  U_1 G_0 |\psi_0\> \\
            &= T^j |\phi_0\>.
\end{align}
In other words, if we find an algorithm whose final state in the case $j=0$ is given by (\ref{eq:finalstates}), then the final state will also be given by (\ref{eq:finalstates}) for $j \in \{1,2,\ldots,N-1\}$.

\subsection{Characterizing algorithms by polynomials}

Another advantage of translation invariant quantum algorithms for the
OSP is that they have a convenient characterization in terms of Laurent
polynomials.  A \emph{Laurent polynomial} is a function $Q:\CC\to\CC$
that can be written as
\begin{align}
  Q(z) = \sum_{i=-D}^{D}q_i z^i
\end{align}
for some positive integer $D$, where each $q_i\in\CC$.  We call $D$
the \emph{degree} of $Q(z)$.  We say $Q(z)$ is \emph{nonnegative}
if, on the unit circle $|z|=1$, $Q(z)$ is real-valued and satisfies
$Q(z) \geq 0$.  Note that for $|z|=1$, $z^* = z^{-1}$, so $Q(z)$ is
real-valued on the unit circle if and only if $q_i = q_{-i}^*$ for
all $i \in \{0,1,\ldots,D\}$.  If $Q(z)=Q(z^{-1})$ for all $z \in \CC$,
\ie, if $q_i=q_{-i}$ for all $i \in \{1,2,\ldots,D\}$, we say $Q(z)$
is \emph{symmetric}.  Thus, $Q(z)$ is nonnegative and symmetric if and only if
$q_i = q_{-i} \in \RR$ for all $z \in \{0,1,\ldots,D\}$.  An example of
a nonnegative, symmetric Laurent polynomial that is relevant to the ordered
search problem is the \emph{Hermite kernel} of degree $N-1$,
\begin{align}
  H_N(z)
  &:= \sum_{i=-(N-1)}^{N-1} 
      \left( 1 - \frac{|i|}{N} \right) z^i \\
  &=  \frac{1}{N} 
      \left( \frac{z^{-N}-1}{z^{-1}-1}\right)
      \left( \frac{z^N-1}{z-1}\right).
\end{align}

The following result of Farhi, Goldstone, Gutmann, and Sipser characterizes
exact translation invariant algorithms for the ordered search problem in
terms of Laurent polynomials.

\begin{theorem}[\cite{Farhi:1999b}]
There exists an exact, translation invariant, $k$-query quantum algorithm
for the $N$-element OSP if and only if there exist nonnegative, symmetric
Laurent polynomials $Q_0(z),\ldots,Q_k(z)$ of degree $N-1$ such that
\begin{align}
  \label{eq:fggspoly1}
  Q_0(z) &= H_N(z) \\
  Q_t(z) &= Q_{t-1}(z) && \text{at~} z^N = (-1)^t \nonumber\\
         &             && \forall\, t\in\{1,2,\ldots,k\} \label{eq:fggspoly2} \\
  Q_k(z) &= 1 \label{eq:fggspoly3} \\
  \frac{1}{2\pi} \int_{0}^{2\pi} \!\! Q_t(e^{i\omega}) \, d\omega &= 1
         && \forall\, t\in\{0,1,\ldots,k\}.
  \label{eq:fggspoly4}
\end{align}
\label{thm:OSPpolys}
\end{theorem}

Each polynomial $Q_t(z)$ in this theorem represents the quantum
state of the algorithm after $t$ queries.  Indeed, if we write
\begin{align}
  Q_t(z) = \sum_{i=-(N-1)}^{N-1} q_i^{(t)} z^i,
\end{align}
then
\begin{align}
  \label{eq:q_il}
  q_i^{(t)} &= 2\sum_{m=1}^{N-i}\<\psi_t|N-m\>\<N-m-i|\psi_t\>,
\end{align}
where
\begin{align}
  |\psi_t\> := U_t G_0 U_{t-1} \ldots U_1 G_0|\psi_0\>
\end{align}
is the state of the quantum computer after $t$ queries when the
target item is $j=0$ \cite{Farhi:1999b}.  Given polynomials satisfying
(\ref{eq:fggspoly1}--\ref{eq:fggspoly4}), one can reconstruct all of
the unitary operators $U_t$ for the algorithm using (\ref{eq:q_il}).

\begin{figure}
\includegraphics[width=\columnwidth]{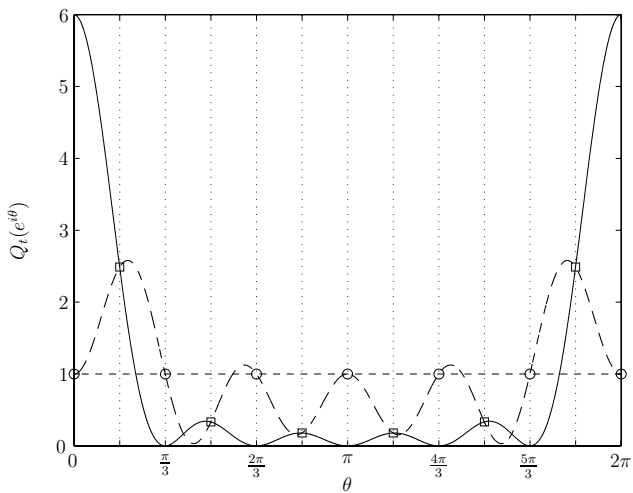}
\caption{$Q_t(e^{i\theta})$ as a function of $\theta$ for $k=2$ and $N=6$.  The solid, long dashed, and short dashed lines represent $t=0$, $1$,
and $2$, respectively.  The intersections at roots of $1$ and $-1$ are indicated by circles and squares, respectively.}
\label{fig:optimalk2poly}
\end{figure}

Figure~\ref{fig:optimalk2poly} shows the (unique) solution to
(\ref{eq:fggspoly1}--\ref{eq:fggspoly4}) for $k=2$ and $N=6$
\cite{Farhi:1999b}.  In general, the polynomial $Q_0(z)$ (the Hermite
kernel) characterizes complete ignorance of the target location at the
beginning of the algorithm, and subsequent polynomials become flatter and
flatter until the final polynomial $Q_k(z)=1$ is reached, corresponding to
exact knowledge of the target location.  Because each query can only change
the quantum state in a restricted way, successive polynomials must agree at
certain roots of $\pm 1$.  Also, each polynomial must be nonnegative and suitably normalized.

With $k=2$, there is a unique choice for the polynomial $Q_1(z)$, which
might or might not be nonnegative depending upon the value of $N$.  For $N
\le 6$, this polynomial is nonnegative (showing that an ordered list of size
$N \le 6$ can be searched in two quantum queries), whereas for $N \ge 7$, it
is not \cite{Farhi:1999b}.

The best ordered search algorithm discovered by Farhi \etal\ was found
by considering $k=3$ queries.  For fixed values of the degree $N-1$,
they numerically searched for polynomials $Q_1(z),Q_2(z)$ satisfying
the constraints (\ref{eq:fggspoly1}--\ref{eq:fggspoly4}) of Theorem
\ref{thm:OSPpolys}.  The largest value of $N$ for which they found a
solution was $N=52$.  Applying this $52$-item ordered search algorithm
recursively gives an algorithm for instances with $N$ arbitrarily
large.  Specifically, one divides the list into $52$ sublists and
applies the algorithm to the largest (rightmost) item of each sublist,
finding the sublist that contains the target in $3$ queries.  This
process repeats, with every $3$ queries dividing the problem size by
$52$, leading to a query complexity of $3\lceil\log_{52} N\rceil$.
(Note that although the base algorithm in this recursion is
translation invariant, the scalable algorithm generated in this way is
not.)

In general, recursion can be used to turn small base cases into
scalable algorithms, so improved quantum algorithms for the OSP can be
found by discovering improved base cases.  Subsequent work by one of
us (AJL) and collaborators sought such algorithms using a conjugate
gradient descent search for the polynomials $Q_t(z)$
\cite{Brookes:2004a}.  This method is guaranteed to work (for a small
enough step size) because the space of polynomials satisfying
(\ref{eq:fggspoly1}--\ref{eq:fggspoly4}) is convex.  The best
solutions found by this method were $N=56$ for $k=3$ and $N=550$ for
$k=4$, implying a $4\log_{550} N \approx 0.439 \log_2 N$ query
recursive algorithm.  Unfortunately, conjugate gradient descent (or
any approach based on local optimization) can never prove that finite
instance algorithms do not exist for a given number of queries $k$.
It could always be the case that lack of progress by a solver is
indicative of inadequacies of the solver (\eg, the step size is too
large, etc.).  In the next section, we recharacterize exact
translation invariant quantum OSP algorithms in a way that allows
either their existence or nonexistence (whichever the case may be) to
be proved efficiently.

\section{A semidefinite program for translation invariant quantum algorithms
for the OSP}
\label{sec:semidefinite}

\subsection{Formulation of the SDP}

In this section, we show that the problem of finding Laurent polynomials
satisfying the conditions of Theorem~\ref{thm:OSPpolys} can be viewed as an
instance of a particular kind of convex optimization problem, namely a {\em
semidefinite program} \cite{Vanderberghe:1996a}.  The basic idea is to use
the spectral factorization of nonnegative Laurent polynomials to rewrite
equations (\ref{eq:fggspoly1}--\ref{eq:fggspoly4}) as linear constraints on
positive semidefinite matrices.

The  spectral factorization of nonnegative Laurent polynomials follows from
the Fej\'{e}r-Riesz theorem:

\begin{theorem}[\cite{Fejer:1915a,Riesz:1915a}]
Let $Q(z)$ be a Laurent polynomial of degree $D$.  Then $Q(z)$
is nonnegative if and only if there exists a polynomial $P(z) =
\sum_{i=0}^{D} p_i z^i$ of degree $D$ such that $Q(z) = P(z) P(1/z^*)^*$.
\end{theorem}

Let $\Tr_i$ denote the trace along the $i$th super-diagonal (or $(-i)$th
sub-diagonal, for $i<0$), \ie, for an $N \times N$ matrix $X$,
\begin{align}
  \Tr_i\, X = \begin{cases} \sum_{\ell=1}^{N-i} X_{\ell,\ell+i} & i \ge 0 \\
                            \sum_{\ell=1}^{N+i} X_{\ell-i,\ell} & i < 0.
              \end{cases}
\end{align}
The Fej\'{e}r-Riesz theorem can be used to express nonnegative Laurent
polynomials in terms of positive semidefinite matrices, as shown by the
following lemma.

\begin{lemma}
Let $Q(z) = \sum_{i=-(N-1)}^{N-1} q_i z^i$ be a Laurent polynomial of degree
$N-1$.  Then $Q(z)$ is nonnegative if and only if there exists an $N\times
N$ Hermitian, positive semidefinite matrix $Q$ such that $q_i = \Tr_i\,Q$.
\label{lem:basicsos}
\end{lemma}

\begin{proof}
The ``if'' direction follows from the representation
\begin{align}
  Q(z) &= 
  \begin{bmatrix} 1 & \cdots & z^{-(N-1)} \end{bmatrix}
  Q
  \begin{bmatrix} 1 \\ \vdots \vspace{.5ex} \\ z^{N-1} \end{bmatrix}.
\end{align}
This $Q(z)$ is real on $|z|=1$ since $Q=Q^\dagger$; it is nonnegative there
because $Q$ is positive semidefinite.

The converse follows from the spectral factorization of $Q(z)$.  Let $Q(z) =
P(z) P(1/z^*)^*$, let $\textbf{p} := \begin{bmatrix}p_0 & \cdots &
p_{N-1}\end{bmatrix}^T$, and let $\textbf{z} := \begin{bmatrix}1 & \cdots &
z^{N-1}\end{bmatrix}^T$. Then $P(z) = \textbf{p}^T \textbf{z}$, and $Q(z) = \textbf{z}^\dagger \textbf{p}^* \textbf{p}^T \textbf{z}$ on $|z|=1$.  We choose $Q := \textbf{p}^* \textbf{p}^T$, which by construction is Hermitian and positive semidefinite.  Furthermore, since $Q(z)$ on $|z|=1$ determines the coefficients $q_i$, we have $q_i = \Tr_i\,Q$.
\end{proof}

Because the Laurent polynomials in Theorem \ref{thm:OSPpolys} are not
only nonnegative but also \emph{symmetric}, we can restrict the
associated matrices to be real symmetric, as the following lemma shows.

\begin{lemma}
\label{lem:realQ}
If $Q(z)$ is a nonnegative, symmetric Laurent polynomial, then the
matrix $Q$ in Lemma~\ref{lem:basicsos} can be chosen to be real and
symmetric without loss of generality.
\end{lemma}
\begin{proof}
Let $Q$ be a Hermitian, positive semidefinite matrix such that $Q(z) = 
\textbf{z}^\dagger Q \textbf{z}$ on $|z|=1$, where $\mathbf{z}$ is defined as in the proof of Lemma~\ref{lem:basicsos}.  Then the symmetry $Q(z) = Q(z^{-1})$
implies that $Q(z) = \textbf{z}^\dagger Q^T \textbf{z}$ on $|z|=1$, and by averaging these two expressions, we have $Q(z) = \textbf{z}^\dagger \tilde Q \textbf{z}$ on $|z|=1$, where $\tilde Q := (Q + Q^T)/2$ is real and symmetric.
\end{proof}

Using Lemma \ref{lem:realQ}, we can recast the conditions
(\ref{eq:fggspoly1}--\ref{eq:fggspoly4}) of Theorem \ref{thm:OSPpolys}
as the following \emph{semidefinite program}:

\begin{sdp}[$S(k,N)$]
\label{sdp:FGGSprimal}
Find real symmetric positive semidefinite $N\times N$ matrices
$Q_0,Q_1,\ldots,Q_k$ satisfying
\begin{align}
  \label{eq:fggssdp1}
  Q_0 &= E/N \\
  \mathcal{T}_t\, Q_t &= \mathcal{T}_t\, Q_{t-1} 
    &&\forall\, t \in \{1,2,\ldots,k\} \label{eq:fggssdp2} \\
  Q_k &= I/N \label{eq:fggssdp3} \\
  \Tr\, Q_t &= 1 &&\forall\, t \in \{0,1,\ldots,k\}
  \label{eq:fggssdp4}
\end{align}
where $E$ is the $N \times N$ matrix in which every element is $1$ and
$\mathcal{T}_t:\mathcal{S}^{N} \to \RR^{N-1}$ is a linear operator
(on the space $\mathcal{S}^{N}$ of real symmetric $N\times N$
matrices) that computes signed traces along the (off-) diagonals, namely
\begin{align}
  (\mathcal{T}_t\, X)_i &:= \Tr_i \,X + (-1)^t \, \Tr_{i-N} \,X
\end{align}
for $i \in \{1,2,\ldots,N-1\}$.
\end{sdp}

The existence of an exact, translation invariant quantum algorithm for
the OSP is equivalent to the existence of a solution to this
semidefinite program, which can be seen as follows:

\begin{theorem}
\label{thm:sdpequivalence}
There exists an exact, translation invariant, $k$-query quantum algorithm
for the $N$-element OSP if and only if $S(k,N)$ has a solution.
\end{theorem}

\begin{proof}
Given $Q_0,Q_1,\ldots,Q_k$ satisfying $S(k,N)$, let $Q_j(z) :=
\begin{bmatrix}1 & \cdots & z^{-(N-1)}\end{bmatrix} Q_j
\begin{bmatrix}1 & \cdots & z^{N-1}\end{bmatrix}^T$.
Then the symmetry of each matrix $Q_j$ implies that each $Q_j(z)$ is a
nonnegative, symmetric Laurent polynomial; and conditions
(\ref{eq:fggssdp1}--\ref{eq:fggssdp4}) imply conditions
(\ref{eq:fggspoly1}--\ref{eq:fggspoly4}), respectively.

Conversely, suppose $Q_0(z),Q_1(z),\ldots,Q_k(z)$ are nonnegative,
symmetric Laurent polynomials of degree $N-1$ satisfying
(\ref{eq:fggspoly1}--\ref{eq:fggspoly4}).  Let $Q_0:=E/N$, let
$Q_k:=I/N$, and let $Q_1,Q_2,\ldots,Q_{k-1}$ be positive semidefinite
matrices obtained from $Q_1(z),Q_2(z),\ldots,Q_{k-1}(z)$ according to
Lemma~\ref{lem:realQ}.  Then (\ref{eq:fggspoly2}) and
(\ref{eq:fggspoly4}) imply (\ref{eq:fggssdp2}) and
(\ref{eq:fggssdp4}), respectively.
\end{proof}

This reformulation of the problem has the advantage that semidefinite
programs are a well-studied class of convex optimization problems.  In
fact, semidefinite programming feasibility problems can be solved
(modulo some minor technicalities) in polynomial time
\cite{Vanderberghe:1996a,Vanderberghe:2004a}.  Furthermore, there are
several widely available software packages for solving semidefinite
programs \cite{Sturm:2001a,Toh:2002a,SDPA:2005a}.

Note that by ``solving'' a semidefinite program, we mean not only finding a
solution if one exists, but also generating an \emph{infeasibility
certificate} (namely, a solution to the dual semidefinite program) if one
does not.  Thus, by solving $S(k,N)$ for various values of $k$ and $N$, not
only can we extract algorithms from feasible solutions, but we can also
generate lower bounds for the quantum query complexity of the OSP (assuming
we restrict our attention to exact, translation invariant algorithms).  In
other words, this approach unifies algorithm design and lower bound analysis
into a single method.

\subsection{Improved formulation by symmetry reduction}
\label{sec:symmetry}

In moving from the polynomial to the semidefinite programming formulation,
we have increased the number of real parameters specifying an exact,
translation invariant quantum OSP algorithm from $(N-1)(k-1)$ to
$N(N+1)(k-1)/2$.  As benefits, we have put the problem in a numerically
tractable form, and we are now able to prove nonexistence as well as
existence of algorithms.  But the increase in parameters is nevertheless
undesirable.

Fortunately, in our case we can reduce the size of the parameter set
roughly by half by exploiting symmetry.
In particular, in terms of the $N \times N$ counterdiagonal matrix (the
\emph{counteridentity matrix})

\begin{align}
  J &:= 
  \begin{bmatrix}
  0      & 0      & \cdots    & 1 \\
  \vdots & \vdots & \revddots & \vdots \\
  0      & 1      & \cdots    & 0 \\
  1      & 0      & \cdots    & 0 
\end{bmatrix},
\end{align}
we have

\begin{lemma}
If $Q_0,Q_1,\ldots,Q_k$ is a solution to $S(k,N)$, then so is
$JQ_0J,JQ_1J,\ldots,JQ_kJ$.
\end{lemma}

\begin{proof}
The matrices $J Q_t J$ are positive semidefinite since $J$ is unitary.
Clearly, $J Q_0 J = Q_0$ and $J Q_k J = Q_k$, so (\ref{eq:fggssdp1}) and
(\ref{eq:fggssdp3}) are satisfied.  Since $\Tr_i \, JQ_tJ = \Tr_{-i}
\, Q_t$ by the definition of $J$, and since $\Tr_{-i} \, Q_t = \Tr_i
\, Q_t$ because each $Q_t$ is a symmetric matrix, (\ref{eq:fggssdp3})
is satisfied.  Finally, (\ref{eq:fggssdp4}) is satisfied since $J^2=I$.
\end{proof}

\noindent
Thus, by convexity, if $Q_0,Q_1,\ldots,Q_k$ is a solution to $S(k,N)$
then so is $\frac{1}{2}(Q_0 + J Q_0 J),\frac{1}{2}(Q_1 + J Q_1
J),\ldots,\frac{1}{2}(Q_k + J Q_k J)$.  In other words, we can assume
that the matrices $Q_t$ commute with $J$ without loss of generality.

We can use group representation theory to harness this symmetry.  Note that
$\{I, J\}$ is an $N$-dimensional (reducible) representation of the group
$\ZZ/2$.  Since $J$ has $\lfloor N/2 \rfloor$ eigenvalues equal to $-1$ and
the rest equal to $+1$, this representation can be diagonalized into
$\lfloor N/2\rfloor$ copies of the sign representation and $\lceil
N/2\rceil$ copies of the trivial representation by some unitary matrix $U$.
The set of matrices that commute with all matrices in this representation
(the representation of the \emph{commutant subalgebra} of $\ZZ/2$) are
therefore block-diagonalizable (by the same matrix $U$) into two blocks,
with one block having size $\lfloor N/2 \rfloor$ and the other having size
$\lceil N/2 \rceil$.  When $N$ is even,
\begin{align}
  U &= \frac{1}{\sqrt{2}}
  \begin{bmatrix}
  I & I \\
  J & -J
  \end{bmatrix}, 
\end{align}
and when $N$ is odd,
\begin{align}
  U &= \frac{1}{\sqrt{2}}
  \begin{bmatrix}
  I & 0 & I \\
  0 & \sqrt{2} & 0 \\
  J & 0 & -J
\end{bmatrix}, 
\end{align}
where $I$ and $J$ are the $\lfloor N/2\rfloor$ by $\lfloor N/2\rfloor$
identity and counteridentity matrices, respectively.

Now, since we can choose the matrices $Q_t$ to commute with $J$ without loss
of generality, we can block-diagonalize them into twice as many matrices,
each of which has about one quarter the number of elements.  For example, for
$N$ even, $Q = JQJ$ implies that $Q$ has the form
\begin{align}
  Q &= 
  \begin{bmatrix}
  A & B \\ JBJ & JAJ
  \end{bmatrix}
\end{align}
where $A = A^T$ and $B = J B^T J$. Thus we have 
\begin{align}
  U^\dag Q U &=
  \frac{1}{2}
  \begin{bmatrix} I & I \\ J & -J  \end{bmatrix}^T  
  \begin{bmatrix} A & B \\ JBJ & JAJ  \end{bmatrix} 
  \begin{bmatrix} I & I \\ J & -J  \end{bmatrix} \\
  &= 
  \begin{bmatrix} A + B J & 0 \\ 0 & A-BJ  \end{bmatrix},
\end{align}
so that $Q$ is positive semidefinite if and only if $A \pm B J$ are both
positive semidefinite.  The net effect of this symmetry reduction is
to cut the number of real parameters in $S(k,N)$ to $N(N/2+1)(k-1)/2$
(for $N$ even) or $(N+1)^2(k-1)/4$ (for $N$ odd), \ie, roughly by half.

\section{Results and discussion}
\label{sec:results}

We solved the semidefinite program $S(k,N)$ for various values of $k$ and
$N$ using the numerical solvers SeDuMi \cite{Sturm:2001a}, SDPT3
\cite{Toh:2002a}, and SDPA \cite{SDPA:2005a}.  These solvers use
general-purpose primal-dual interior-point methods that eventually become
limited by machine memory.  (Although there are algorithms for solving
SDPs that are not based on interior point methods, we did not attempt to use
such algorithms.)

The time required to solve $S(k,N)$ was substantially reduced by exploiting
the symmetry described in Section~\ref{sec:symmetry}.  In addition, it is
helpful that the constraints are fairly sparse.   Nevertheless, we are
ultimately limited by the fact that the maximum size of a list that can be
searched increases exponentially with the number of queries, so that we can
only consider fairly small values of $k$.

For each $k \leq 4$, we found the smallest value $N^*$ such that
$S(k,N^*)$ has a solution but $S(k,N^*+1)$ does not.  Although we were
able to find solutions to $S(5,N)$ for some values of $N$, we ran out
of machine memory before we could find an infeasibility certificate.
A summary of the values $N^*$ we obtained is presented in
Table~\ref{tab:results}.

\begin{table}
\begin{center}
\begin{tabular}{c@{~}|@{~}c}
  $k$ & $N^*$ \\ \hline
  2   & 6     \\
  3   & 56    \\
  4   & 605   \\
  5   & $>$ 5000
\end{tabular}
\end{center}
\caption{Ordered list sizes $N^*$ searchable a $k$-query exact,
translation invariant quantum algorithm such that no such algorithm
exists for a list of size $N^*+1$.}
\label{tab:results}
\end{table}

\begin{figure}
\includegraphics[width=\columnwidth]{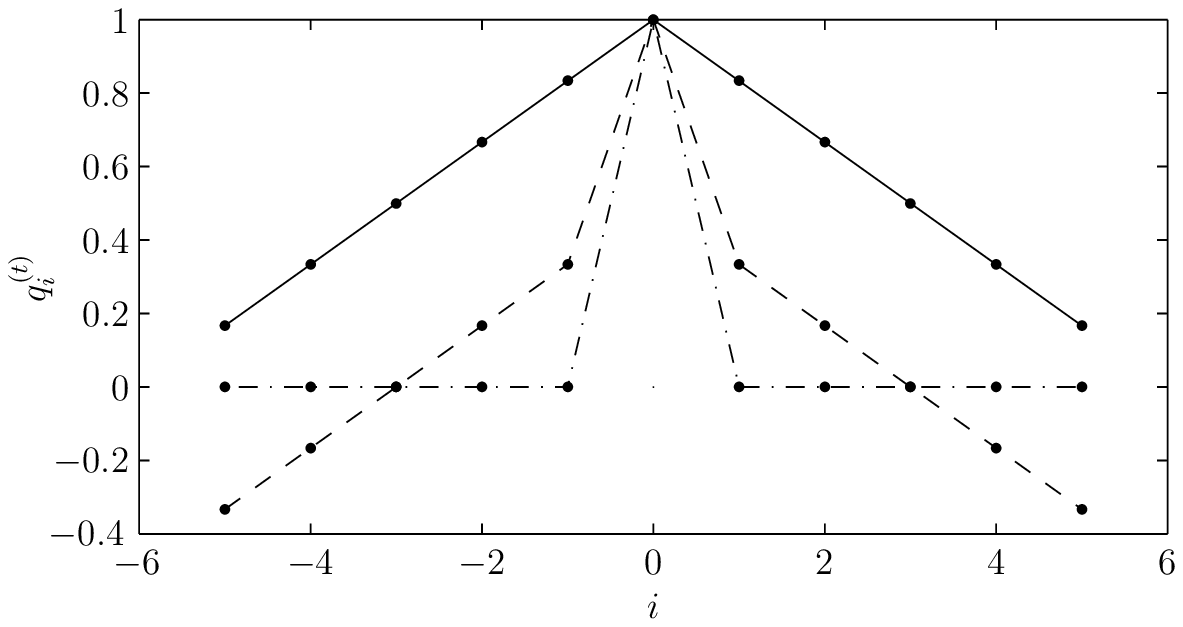}
\includegraphics[width=\columnwidth]{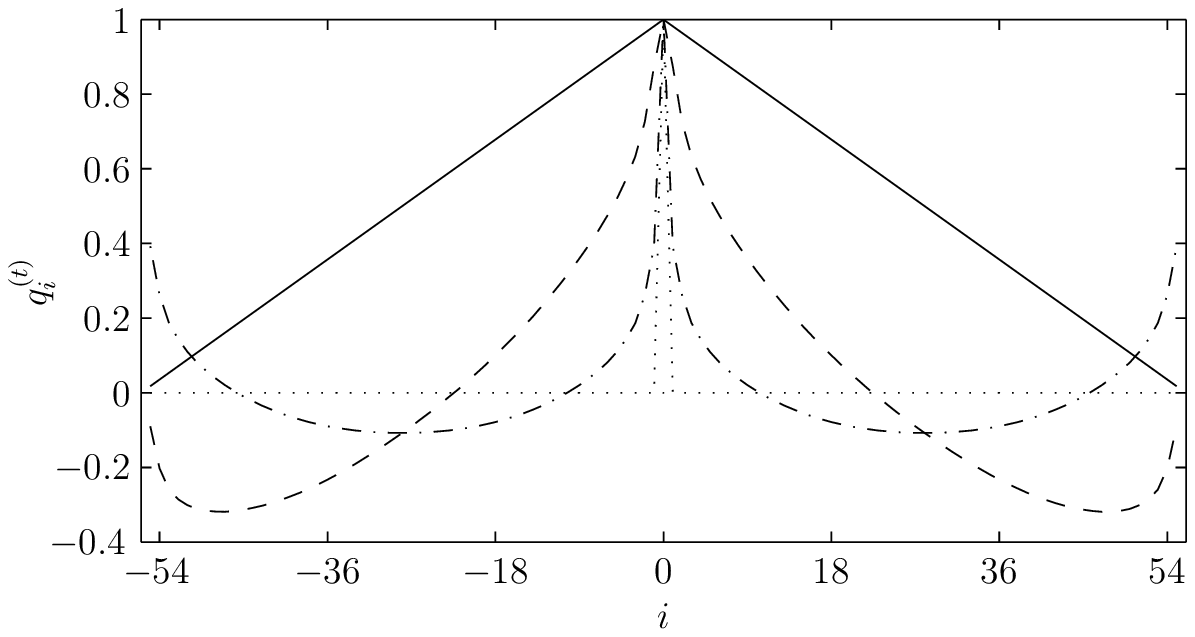}
\includegraphics[width=\columnwidth]{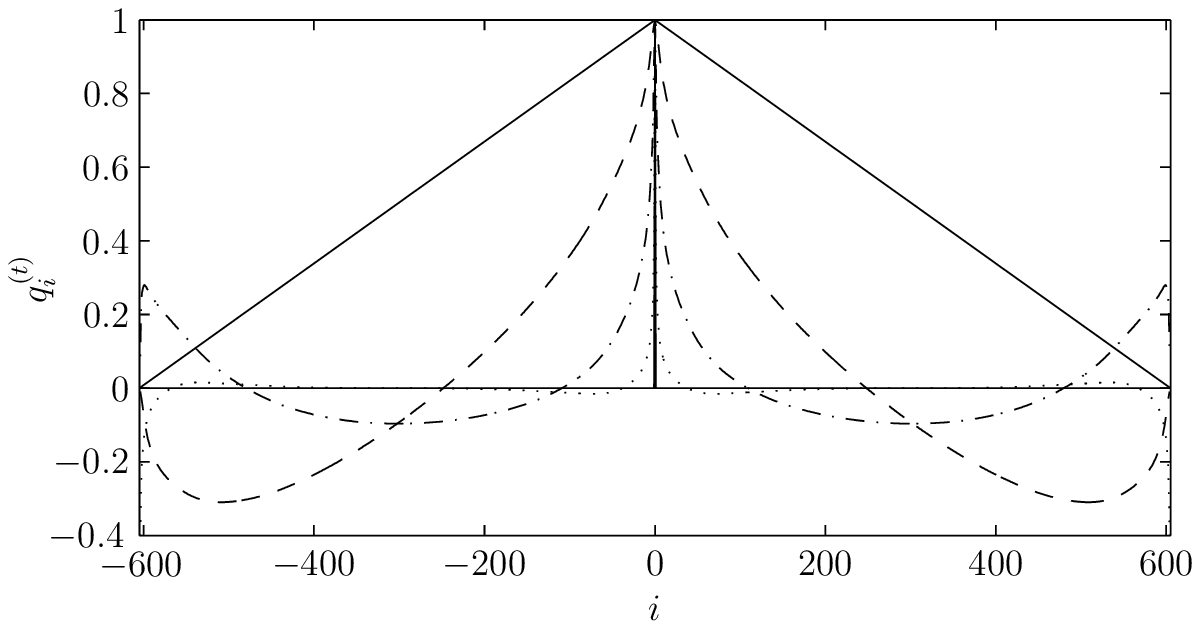}
\caption{Laurent polynomial coefficients $q_i^{(t)}$ as a function of $i$ for exact, translation invariant OSP algorithms.  From top to bottom, $k=2$, $3$, and $4$, with $N=6$, $56$, and $605$, respectively.}
\label{fig:optimalcoeffs}
\end{figure}

By recursion, the $k=4$, $N^*=605$ query algorithm yields a scalable
algorithm whose query complexity is
\begin{align}
  4\log_{605} N \approx 0.433 \log_2 N.
\end{align}
This result also implies improvements to other algorithms; for example, it implies a quantum sorting algorithm whose query complexity is $4N\log_{605} N$.

As mentioned in the introduction, infeasibility of $S(k,N^*+1)$ does not necessarily imply that $N^*$ is the largest size of a list that can be searched with a $k$-query exact, translation invariant algorithm.  However, it seems reasonable to conjecture that this might be the case.  Indeed, for $k=2$ and $3$, we know that the values of $N^*$ in Table~\ref{tab:results} are  optimal.  For those smaller problems, we were able to numerically solve the larger SDP developed by Barnum, Saks, and Szegedy to characterize general quantum query algorithms \cite{Barnum:2003a}.  (To solve large enough instances of that SDP, it was also crucial for us to exploit the symmetry of the problem.)  Those results show that $N=6$ and $N=56$ are the largest sizes of lists that can be searched with $k=2$ and $k=3$ queries, respectively, even when the assumption of translation invariance is removed.

Whether the $\tfrac{1}{\pi}\ln N$ lower bound on the query complexity of the
OSP can be saturated remains open.  However, the structure of the algorithms
we obtained suggests the possibility of a well-behaved analytic solution,
and it would be interesting to understand the behavior of the solution in
the limit of large $N$.  Figure~\ref{fig:optimalcoeffs} shows the
coefficients of the polynomials $Q_t(z)$ associated with the optimal
feasible solutions $Q_t$ for $k=2,3,4$.  Note the similarity of the
coefficients for different values of $N$.

Not only have we found a particular quantum algorithm for the ordered
search problem, but we have also demonstrated the usefulness of
semidefinite programming as a numerical technique for discovering
quantum query algorithms.  Indeed, the connection between quantum
query complexity and convex optimization is not unique to the ordered
search problem: as mentioned above, arbitrary quantum query problems
can be characterized in terms of semidefinite programs
\cite{Barnum:2003a}. Thus, semidefinite programming appears to be a
powerful tool for studying quantum query complexity.

After this work was completed, we learned that Ben-Or and Hassidim have developed a new approach to quantum algorithms for ordered search based on adaptive learning \cite{BH06}.  Their resulting algorithm is not exact, but rather, is zero-error, with a stochastic running time (sometimes referred to as a Las Vegas algorithm).  The expected running time of their algorithm is $0.32 \log_2 N$.


\begin{acknowledgments}
AMC is supported in part by the National Science Foundation under
contract number PHY-0456720.
AJL is supported in part by Army Research Office contract number
W911NF-04-1-0242.
PAP is supported in part by AFOSR MURI subaward 2003-07688-1 and the
Singapore-MIT Alliance.
While previously at the MIT Center for Theoretical Physics, AMC and AJL
received support from the Department of Energy under cooperative research
agreement DE-FC02-94ER40818 and from the Cambridge-MIT Institute, AMC
received support from the Fannie and John Hertz Foundation and from the
National Security Agency and Advanced Research and Development Activity
under Army Research Office contract DAAD19-01-1-0656, and AJL received support
from a Hewlett Packard-MIT postdoctoral fellowship.

\end{acknowledgments}


\end{document}